\documentclass[aps,prl,reprint,twocolumn]{revtex4-2}
\usepackage[usenames,dvipsnames]{xcolor}
\usepackage{lipsum}
\usepackage{graphicx}
\usepackage{array}
\usepackage{epstopdf}
\usepackage[justification=justified]{caption}
\usepackage{amsmath}
\usepackage{multirow}
\usepackage{amsmath}
\usepackage{amsfonts}
\usepackage{amssymb}
\usepackage{mathrsfs}
\usepackage{mathtools}
\usepackage{tikz}
\usepackage{pgfplots}
\usepackage{csquotes}

\definecolor{color1}{RGB}{0,113,188}
\definecolor{color2}{RGB}{216,82,24}
\definecolor{color3}{RGB}{236,176,31}

\begin{document}

\author{Igor A. Maia}
\author{Peter Jordan}
\affiliation{Pprime Institute, CNRS, Universit\'e de Poitiers, ENSMA, Poitiers, France}
\author{Andr\'e V. G. Cavalieri}
\author{Eduardo Martini}
\affiliation{Divis\~ao de Engenharia Aeron\'autica, Instituto Tecnol\'ogico de Aeron\'autica, S\~ao Jos\'e dos Campos, Brazil}
\author{Fl\'avio J. Silvestre}
\affiliation{Institute of Aeronautics and Astronautics, Technische Universit\"at Berlin, Germany}

\title{Closed-loop control of forced turbulent jets}

\begin{abstract}


Closed-loop control of turbulent flows is a challenging problem with important practical and fundamental implications. We perform closed-loop control of forced, turbulent jets based on a wave-cancellation strategy. The study is motivated by the success of recent studies in applying wave cancellation to control instability waves in transitional boundary layers and free-shear flows. Using a control law obtained through a system-identification technique, we successfully implement wave-cancellation-based, closed-loop control, achieving order-of-magnitude attenuations of velocity fluctuations. Control is shown to reduce fluctuation levels over an extensive streamwise range. 

\end{abstract}

\maketitle

Flow control, in open or closed-loop, is often applied to flows with low or moderate Reynolds numbers, where the aim is to avoid transition to turbulence and maintain the laminar state \cite{FranssonPRL2006,FabbianeJFM2015}. This is accomplished in situations where the laminar solution is linearly stable \cite{kuhnen_nature2018}, and/or with mild amplification of incoming disturbances \cite{FabbianeJFM2015}. Jets with high Reynolds numbers strongly amplify upstream disturbances through a process associated with convective non-normality \cite{CossuPRL1997}. In such cases, it is unlikely that the laminar solution may be recovered by closed-loop control. An alternative is the application of closed-loop control to stabilise a base state other than the laminar solution \cite{Wang_gibson_waleffe_prl2007}, but the feasibility of such approaches for high-Reynolds-number flows remains to be assessed. 

We here follow a different path aiming to cancel the convective amplification of some disturbances in the jet by an appropriate choice of control action. The targeted disturbances are axisymmetric wavepackets present in the jet. These are known to play a key role in jet-noise, and to be underpinned by dynamics that can be described using linear models \cite{JordanColoniusReview, CavalieriatalAMR2019}. Attenuating axisymmetric wavepackets does not eliminate jet turbulence, which is dominated by other energy-containing structures \cite{CoherenceVincent}, but it has potential to reduce jet noise. Linear stability theory then appears as a candidate to provide suitable control-law design \cite{CavalieriatalAMR2019}. Alternatively, control laws can be obtained by means of transfer functions identified empirically by measuring the flow response to external forces \cite{HerveetalJFM_ARMAX, SasakiJFM2017}.


In a linear scenario, it is possible to devise an actuation that eliminates disturbances through a simple superposition of waves in a destructive pattern, provided that the disturbance is identified by a sensor and there is a suitable model for its space-time evolution. Such wave cancellation has been successfully applied to the control of Tollmien-Schlichting waves in laminar boundary layers \cite{Thomas1983, Li&Gaster2006,SasakiTCFD2018_2}. More recently, \citet{SasakiTCFD2018_1} and \citet{SasakiTCFD2018_2} have associated linear theory, system identification and wave cancellation in numerical studies to perform closed-loop control of laminar boundary layers and transitional mixing layers, respectively. Particularly interesting are the results of \citet{SasakiTCFD2018_2}, which showed that optimal linear control theory produces a similar control law and a comparable performance to the simpler, transfer-function-based, wave-cancellation approach, suggesting that the former is underpinned by the latter.

Such results are encouraging vis-\`a-vis the turbulent jet problem, where coherent structures in the initial region are underpinned by linear mechanisms \cite{JordanColoniusReview}. In an open-loop configuration, \citet{KopievAcPhy2013} have shown wave cancellation to be possible for turbulent, harmonically forced jets. However, the possibility of performing broadband wave cancellation in a fully turbulent jet in closed-loop configuration has not, to date, been assessed.

We tackle this problem in an experiment designed to perform real-time closed-loop control of forced jets. The challenge of controlling high-Reynolds-number jet turbulence experimentally is considerable, and thus we have restricted our analysis to a forced flow, where broadband axisymmetric forcing is artificially introduced at the nozzle lip. The goal of the forcing is to increase wavepacket amplitudes, in the spirit of what was done by \citet{CrowChampagne} and \citet{Moore1977}, making them easier to identify and control. Unforced jets have an energy content that is spread across a broad range of azimuthal wavenumbers, and would thus require a high number of sensors in order to measure high-order azimuthal modes without spatial aliasing. In this sense, forcing the axisymmetric mode allows us to greatly simplify the sensor configuration, which involves an array of microphones in the irrotational nearfield and a target hot wire on the jet centerline.
 
\begin{figure}[!ht]
\centering
\includegraphics[trim=5cm 5.5cm 5cm 7cm, clip=true,width=0.99\linewidth]{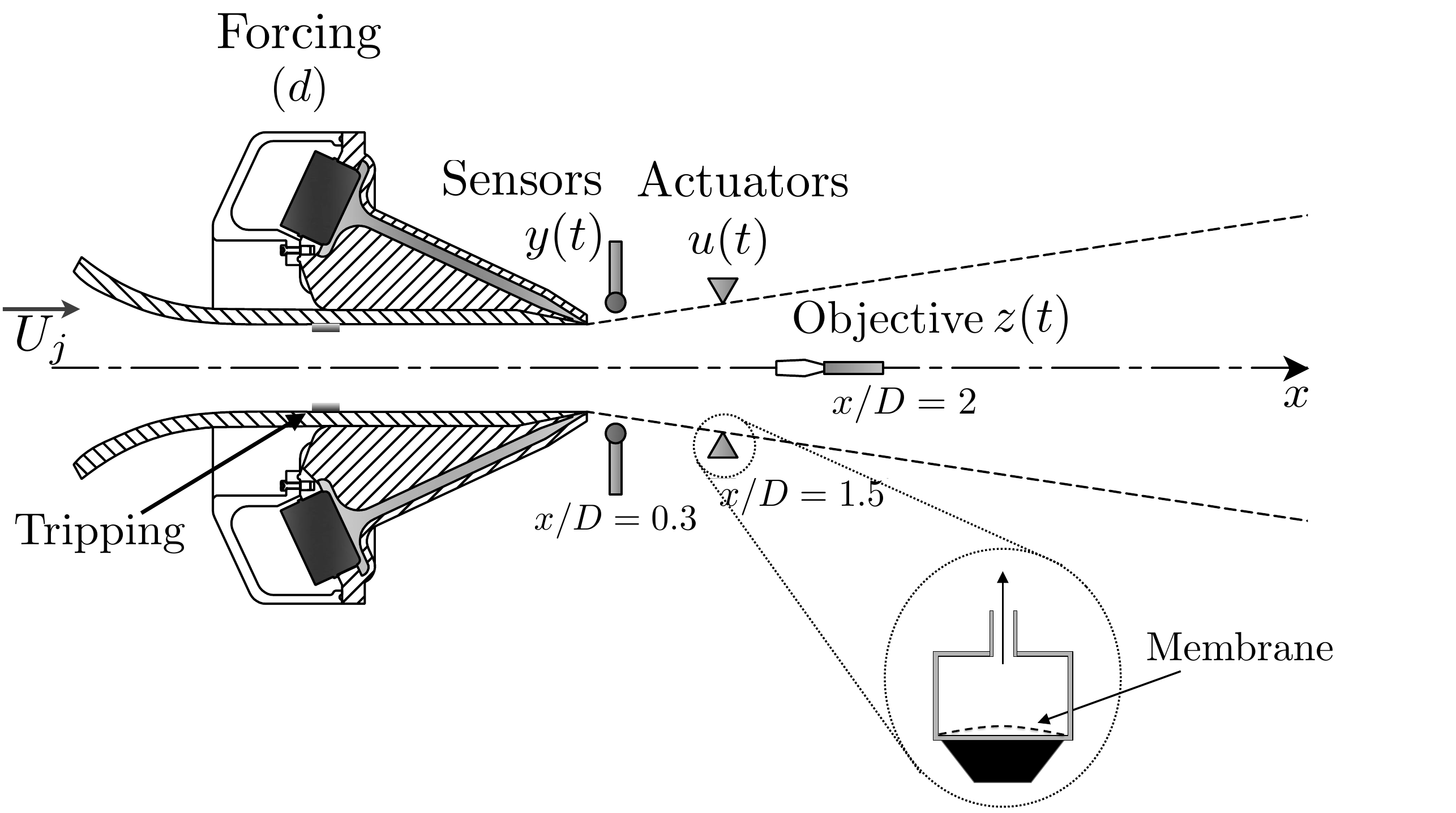}
\caption{\small Schematic of closed-loop control experiment showing the relative position of inputs and outputs. Forcing (at the nozzle lip) and actuation consist of synthetic jets generated by loudspeakers; the sensors are microphones situated in the near-pressure field of the jet, immediately outside of the shear-layer; the objective consists of streamwise velocity measurements performed by a hot wire at the jet centerline. The boundary layer is tripped inside the nozzle by a strip of carborundum particles placed 2.5 diameters upstream of the exit plane, so as to produce a fully turbulent jet.}
\label{setup}
\end{figure}

The experiments were carried out at the Pprime Institute, in Poitiers, France. The jet Mach ($Ma=U_{j}/c_{\infty}$) and Reynolds ($Re=U_{j}D/\nu$) numbers are $0.05$ and $5 \times 10^4$, respectively, with $U_{j}$ the jet exit velocity, $c_{\infty}$ the ambient speed of sound, $D$ the jet diameter and $\nu$ the kinematic viscosity of the air. The nozzle boundary layer is tripped $2.5D$ upstream of the exit plane, to make sure the jet is fully turbulent. Tripping produces a boundary layer at the nozzle exit plane whose velocity profile agrees well with a power law typical of fully developed turbulence. Just donwstream of the exit plane, where control is performed, peak turbulence intensity levels on the shear layer were found to be between 12\% and 15\%. Boundary layer profiles and an aerodynamic characterisation of the jet are reported in the Supplemental Material. The setup consists of four elements: forcing, sensors, actuators and objective (Figure \ref{setup}). The forcing, $d$, consisted of synthetic jets generated by a system of eight loudspeakers (model AURA NSW 2-236-8AT) equally distributed in the azimuthal direction and mounted on a conical structure fitted on the nozzle. The speakers operate in phase, so as to force axisymmetric disturbances. The synthetic jets exit through a $0.01D$ annular gap force the main jet at the nozzle lip. We target axisymmetric, hydrodynamic disturbances because of their importance for jet noise \cite{JordanColoniusReview}. A ring of six $1/4$-inch microphones is placed in the near pressure field at a streamwise position of $0.3D$ from the nozzle exit. The axisymmetric pressure mode measured by the microphones gives the input signal, $y$, for the control law. The actuation, $u$, also consists of synthetic jets generated by synchronised loudspeakers. Six AURA 1-205-8 A speakers are used to drive synthetic jets on a ring array placed immediately outside of the shear-layer at a streamwise position of $1.5D$. The speakers are place inside cavities whose apertures point towards the center of the main jet. Finally, the objective, $z$, consists of streamwise velocity measurements performed using a hot wire situated on the jet centerline downstream of the actuators, at $x/D=2$.  

\begin{figure*}[!ht]
\centering
\includegraphics[trim=0cm 4.5cm 0cm 4.5cm, clip=true, width=0.85\textwidth]{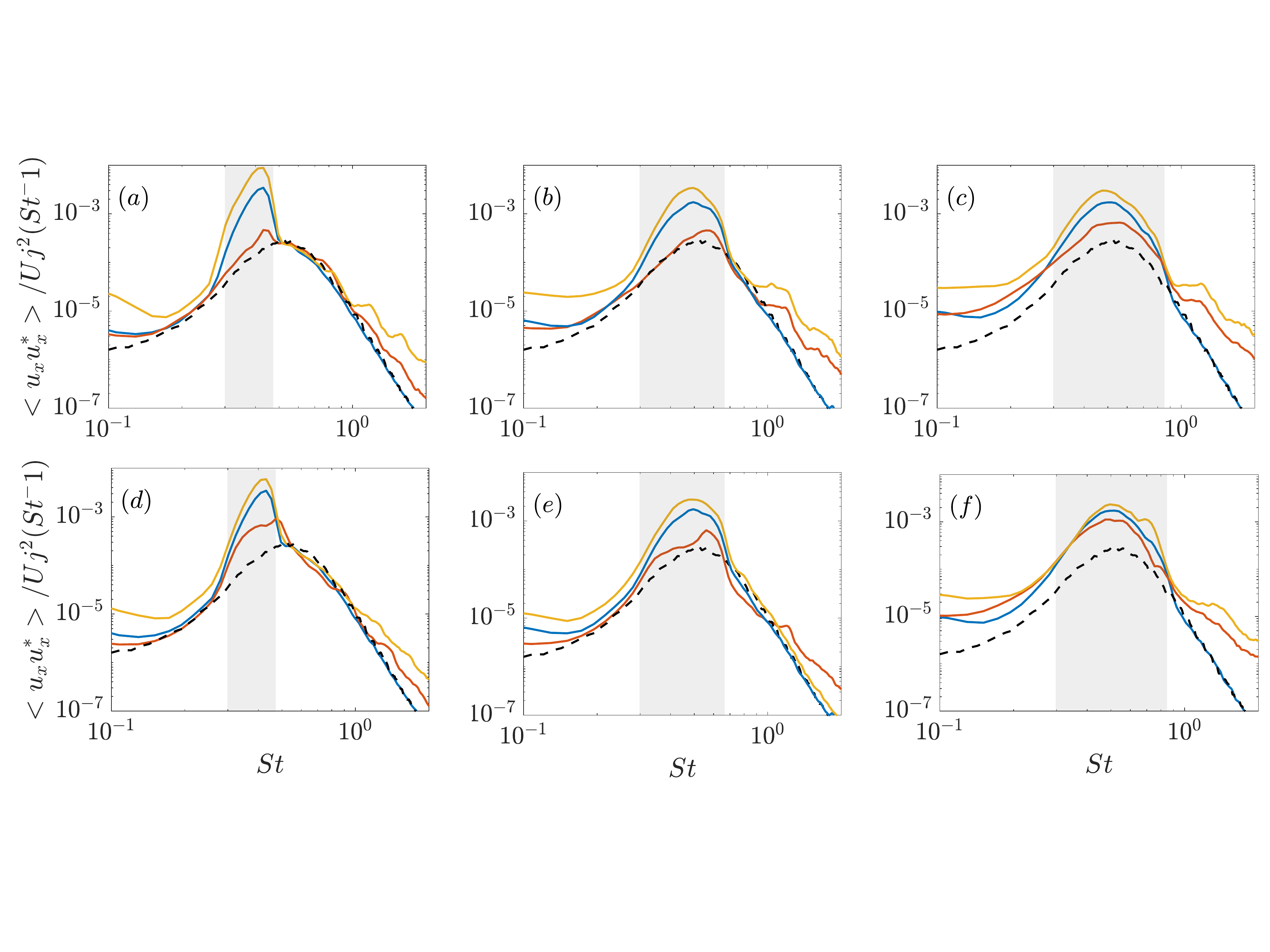}
\caption{ \small Power spectral densities of streamwise velocity fluctuations, $u_{x}$, measured at the objective position ($x/D=2$ at the centerline) of controlled and uncontrolled jets forced at three different bandwidths. \ref{hwplot1}: Baseline case (forced jet); \ref{hwplot2}: controlled jet with `reduction' kernel, $K_{y,d}^{r}$; \ref{hwplot3}: controlled jet with `amplification' kernel, $K_{y,d}^{a}$; \ref{hwplot4}: unforced jet.  (a), (b) and (c): control based on the external disturbances, $d$, as input to the control law; (d), (e) and (f): control based on flow measurements, $y$, as input to the control law. Forcing bandwidths, represented by the grey shaded areas, are: $0.3 \leqslant St \leqslant 0.45$, $0.3 \leqslant St \leqslant 0.65$ and $0.3 \leqslant St \leqslant 0.85$.}
\label{control_res}
\end{figure*}

The control law design is based on that of \citet{SasakiTCFD2018_1,SasakiTCFD2018_2}. We seek to eliminate $z$, which is given as a linear combination of sensor measurements, $y$, plus the actuation signal, $u$. In the frequency-domain, this can be written as:

\begin{equation}
Z(\omega)=Y(\omega)H_{yz}+ U(\omega)H_{uz},
\label{eq1}
\end{equation}

\begin{equation}
U(\omega)=K(\omega)\left(Y(\omega)+H_{uy}U(\omega)\right),
\label{eq2}
\end{equation}
where $Z$, $Y$ and $U$ are the frequency-domain counterparts of $z$, $y$ and $u$. Here, $Z(\omega)$ represents only the part of the measurements that can be estimated from $Y$ and $U$. $H_{yz}$ and $H_{uz}$ are the sensor/objective and actuator/objective transfer functions, respectively, and $H_{uy}$ is a feedback transfer function that accounts for the effect of actuation, $u$, on the measured input, $y$. A transfer function $H_{ij}$ is computed as the ratio between the cross-spectral density (CSD) of $i$ and $j$ and the power-spectral density (PSD) of $i$. The expression for the control kernel, $K$, in the presence of feedback, is given by \cite{SasakiTCFD2018_1,SasakiTCFD2018_2},

\begin{equation}
K_{y}(\omega)=-\frac{H_{yz}(H_{uz}-H_{uy}H_{yz})^{*}}{(H_{uz}-H_{uy}H_{yz})(H_{uz}-H_{uy}H_{yz})^{*}+R},
\label{kernel_y}
\end{equation}
where $^{*}$ denotes a complex conjugate. Here, $R$ is a real-valued penalisation term. As pointed out by \citet{SasakiTCFD2018_2}, the lack of penalisation can lead to a noisy kernel at frequencies for which $H_{yz}$ and $H_{uz}$ have low amplitudes. In order to circumvent this issue, $R$ is adjusted so as to reduce the gain at those frequencies and avoid the appearance of uncontrollable disturbances. In the time domain, the control law is given by,

\begin{equation}
u_{y}(t)=\int_{0}^{\infty}k(\tau)y(t-\tau)\mathrm{d}\tau,
\label{eq3}
\end{equation}
where $k$ is the inverse Fourier transform of $K$. 

We also consider a simplified control problem in which we eliminate the intermediary step of measuring the disturbances upstream of the objective position. This is done by expressing the output as a linear combination of the introduced disturbances and actuation only, leading to the simplified control law,

\begin{equation}
K_{d}(\omega)=-\frac{H_{dz}H_{uz}^{*}}{H_{uz}H_{uz}^{*}+R},
\label{K_ctrl_d}
\end{equation}

\begin{equation}
u_{d}(t)=\int_{0}^{\infty}k(\tau)d(t-\tau)\mathrm{d}\tau.
\label{u_ctrl_d}
\end{equation}

\noindent $d$ acts then at the same time as an external disturbance and an input for the controller. This can be considered as a preliminary trial approach for closed-loop control. It allows us to ascertain whether wave-cancellation is possible in a turbulent jet using closed-loop control. The results of this simplified approach can be considered as the best-case scenario for closed-loop control, because all of the available information about the disturbances is observable and taken into account by the control law. The results so obtained can then be compared to the closed-loop control based on flow measurements.

The closed-loop experiment is carried out using a LabVIEW software. The task of the software is to carry out the convolutions given by Equations \ref{eq3} and \ref{u_ctrl_d} in real-time, using unsteady signals from $y$ or $d$, respectively, as input.

Linearity is a key feature for wave cancellation as is implicit in the linear superposition of Equation \ref{eq1}. Here we use two-point coherence as a measure of the linearity of the system and a criterion for interpretation of the results. Control performance is underpinned by two kinds of coherence: the coherence between disturbance and objective, $\gamma_{\scriptscriptstyle dz}$, or between sensor and objective, $\gamma_{\scriptscriptstyle yz}$ (depending on whether $d$ or $y$ are used as input to the control law), which dictate the accuracy of the estimate of the downstream evolution of disturbances; and the coherence between actuator and objective, $\gamma_{\scriptscriptstyle uz}$, which determines the accuracy of wavepacket generation by the actuators.

A careful preliminary study was carried out in open-loop in order to determine the amplitude of the forcing used in the closed-loop experiment. The amplitudes are selected so as to ensure that the jet response to forcing  falls within a linear regime.

The jet was forced with band-limited white noise signals filtered in three different frequency bands: $0.3 \leqslant St\leqslant 0.45$,  $0.3 \leqslant St \leqslant 0.65$ and $0.3 \leqslant St \leqslant 0.85$, where $St$ is the Strouhal number, given by $St=fD/U_{j}$, with $f$ the frequency. Such forcing produces stochastic phases and amplitudes in the jet response, as opposed to harmonic forcing, as shown in the Supplemental Material. The transfer functions are identified empirically \cite{SasakiJFM2017}, by measuring the response of the jet to forcing and actuation separately. For the gain to be consistent with the disturbances one wishes to control, disturbance/objective, sensor/objective and actuator/objective transfer functions should have the same frequency content. The actuation transfer functions were computed using two kinds of disturbance signals: white-noise and sine sweep, both band-pass filtered in the frequency ranges of interest, and the results obtained were insensitive to the choice of signal. 

Figure \ref{control_res} shows PSDs of streamwise velocity fluctuations, $u_{x}$, measured in controlled and uncontrolled jets at the objective position. The results shown in \ref{control_res}(a), \ref{control_res}(b) and \ref{control_res}(c) were obtained using $d$ as input whereas those in \ref{control_res}(d), \ref{control_res}(e) and \ref{control_res}(f) were obtained using $y$ as input. The spectra of the unforced jet is also shown for comparison. Two kinds of actuation were carried out: one whose gain was designed to reduce disturbance amplitudes, denoted $K_{y,d}^{r}$ (computed through equations \ref{eq3} and \ref{u_ctrl_d}) and another designed to amplify disturbances, $K_{y,d}^{a}$, obtained by applying a $\pi$ phase shift to $K_{y,d}^{r}$. The superscripts $r$ and $a$ denote the application of the reduction- and amplification-aimed kernels to the $y$-based and $d$-based control methods, respectively. 


For the jet forced at the two narrowest frequency bands, control using the two methods ($d$-based and $y$-based) is effective in both reducing and amplifying the disturbances, demonstrating real-time control authority. It is also clear that the $d$-based control performs better than the $y$-based control. Indeed, in the $K_{r}$ configuration the disturbances introduced in the jet are almost entirely eliminated. In the largest frequency band of forcing,  $0.3 \leqslant St \leqslant 0.85$, control performance is degraded for both cases, and amplitudes could not be reduced to the unforced jet levels. Nonetheless, significant reductions are observed.

\scalebox{0}{%
\begin{tikzpicture}
    \begin{axis}[hide axis]
        \addplot [
        color=color1,
        solid,
        line width=1.2pt,
        forget plot
        ]
        (0,0);\label{hwplot1}
        \addplot [
        color=color2,
        solid,
        line width=1.2pt,
        forget plot
        ]
        (0,0);\label{hwplot2}
        \addplot [
        color=color3,
        solid,
        line width=1.2pt,
        forget plot
        ]
        (0,0);\label{hwplot3}
        \addplot [
        color=black,
        dashed,
        line width=1.2pt,
        forget plot
        ]
        (0,0);\label{hwplot4}
    \end{axis}
\end{tikzpicture}%
}%

\begin{figure}[!ht]
\centering
\includegraphics[trim=3cm 3.5cm 4.5cm 3.5cm, clip=true, width=1\linewidth]{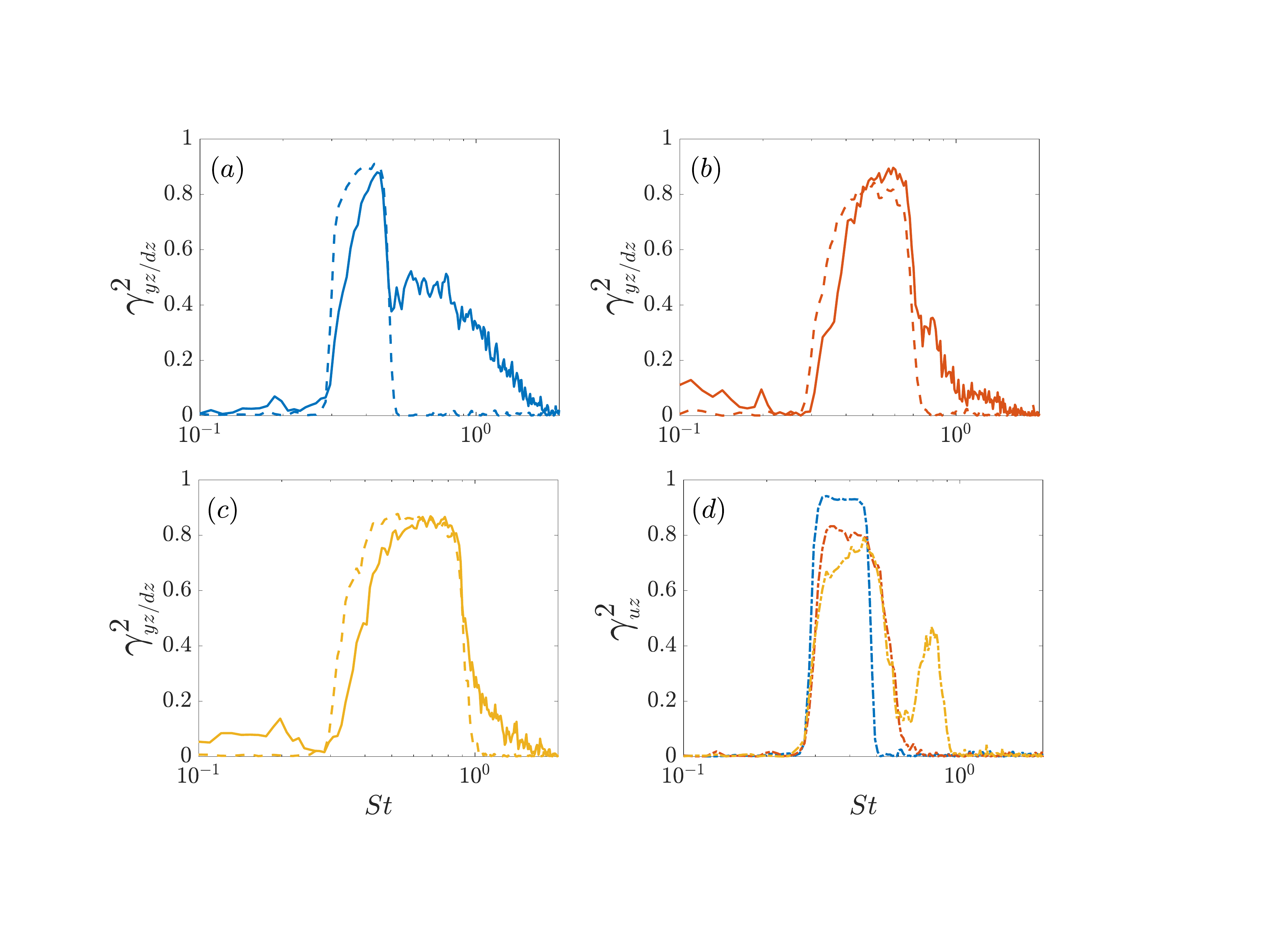}
\caption{\small Coherences associated with the control results shown in Figure \ref{control_res}. (a)-(c): comparison between sensor/objective (solid line), $\gamma_{\scriptscriptstyle yz}$, and disturbance/objective (dashed line), $\gamma_{\scriptscriptstyle dz}$, coherences measured with increasing frequency bandwidth. (d): actuator/objective coherences, $\gamma_{\scriptscriptstyle uz}$, measured for different actuation frequency bands. Forcing bandwidths are: \ref{hwplot1}: $0.3\leqslant St \leqslant 0.45$; \ref{hwplot2}: $0.3\leqslant St \leqslant 0.65$;\ref{hwplot3}: $0.3\leqslant St \leqslant 0.85$. The coherence underpins the effectiveness of the control law, and the drop in $\gamma_{uz}$ with increasing actuation bandwidth is responsible for the degradation of control performance seen in Figure \ref{control_res}.}
\label{coherences}
\end{figure}

\begin{figure}[!ht]
\centering
\includegraphics[trim=3cm 0.9cm 5cm 0.5cm, clip=true, width=1\linewidth]{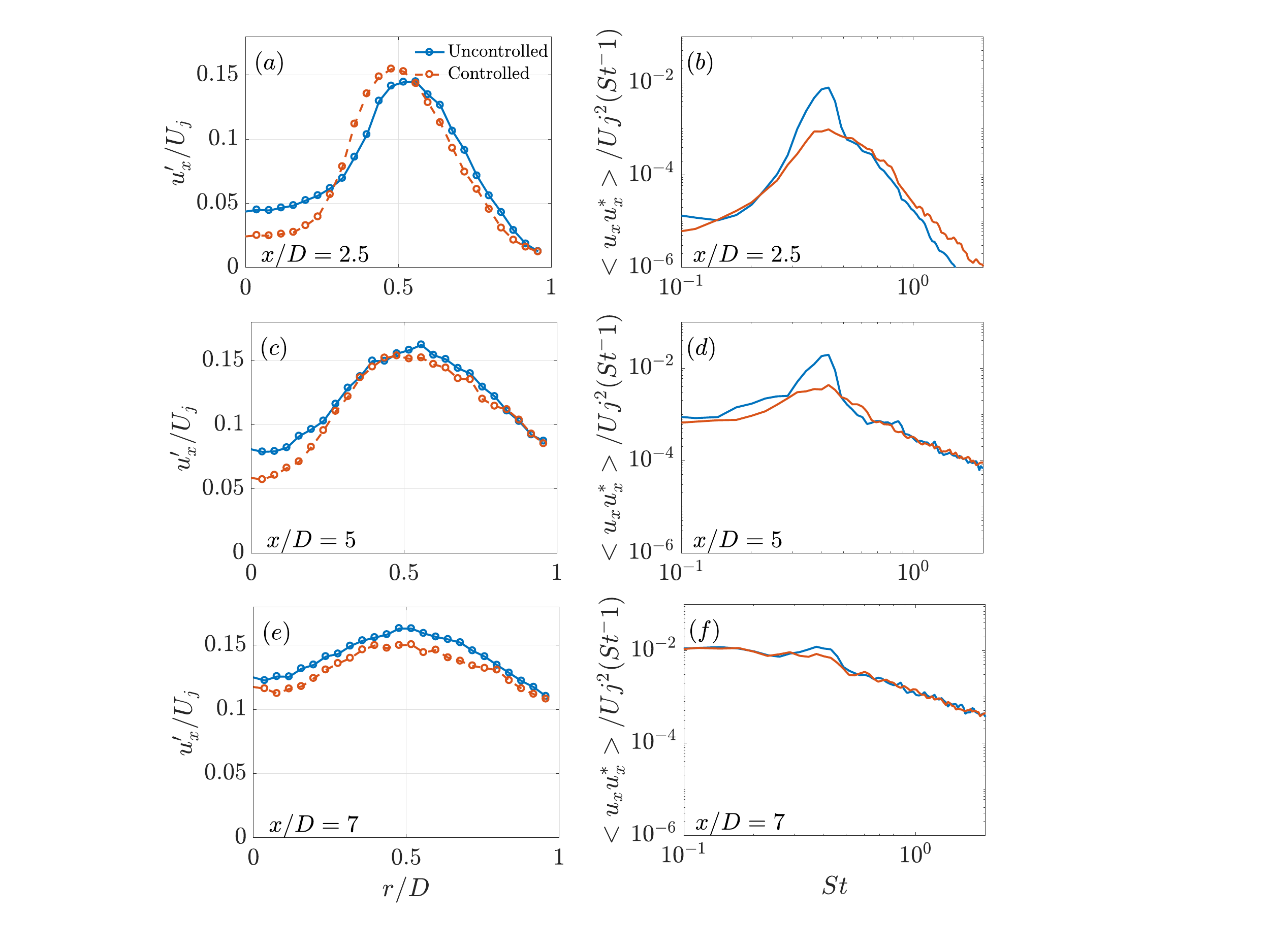}
\caption{ \small Comparison between controlled and uncontrolled jets downstream of the objective position.  (a), (c) and (e): radial profiles of streamwise rms velocity, obtained by integration of the frequency spectrum; (b), (d) and (f): Power spectral densities of streamwise velocity fluctuation measured at the jet centerline. The uncontrolled case corresponds to the baseline jet, forced in the bandwidth $0.3 \leqslant St \leqslant 0.45$, and the controlled case was obtained with a `reduction' kernel, $K_{d}^{r}$. The results show the persistence of control effects as far downstream as 7 diameters.}
\label{effect_down}
\end{figure}

These trends can be understood in light of two-point coherences associated with the transfer functions. The control law is underpinned by estimation and actuation. In the estimation step, the downstream evolution of the disturbances is predicted as they reach the objective position. In this case, $\gamma_{\scriptscriptstyle dz}$ and $\gamma_{\scriptscriptstyle yz}$ are the parameters that determine the accuracy of the estimates. In the actuation step, the incoming wavepackets are eliminated by wavepackets excited by the actuator with the correct phase and amplitude; the accuracy of this step is dictated by the values of $\gamma_{\scriptscriptstyle uz}$. Coherence values close to unity indicate a quasi-linear behaviour, which leads to accurate transfer functions; coherence loss, on the other hand, is associated with nonlinearity \cite{CoherenceJFMAndre, CavalieriatalAMR2019}, and may result in poorly estimated objectives.

Figure \ref{coherences} shows the behaviour of the three important types of coherences as a function of $St$ for the jet forced in different frequency bands. 
 Within each frequency band, $\gamma_{\scriptscriptstyle dz} \geq \gamma_{\scriptscriptstyle yz}$. This results in a better estimation of the forced disturbances, which partially explains the superior performance in comparison with the $y$-based control.


We also observe in \ref{coherences}(d) that, with increasing frequency bands of forcing, there is a severe drop in $\gamma_{\scriptscriptstyle uz}$. This may be associated with two issues: the first is that the jet response to actuation may be nonlinear; the second issue is that actuators are placed outside of the region of non-zero mean flow. Therefore, in order to produce an actuation signal with amplitudes sufficient to eliminate the disturbances introduced upstream, one is obliged to increase amplitude past the linear zone, triggering nonlinear actuator behaviour. Regardless of the precise cause, coherence loss due to nonlinearity becomes more prominent at the higher frequency band of actuation and leads to the degradation of control performance seen in Figures \ref{control_res}(c) and \ref{control_res}(f).

We also investigated the effect of the control on the downstream evolution of the forced wavepackets beyond the objective position. Figure \ref{effect_down} shows streamwise velocity spectra at the jet centerline and radial profiles of streamwise rms velocity, $u_{x}'$, of uncontrolled and controlled jets, measured at three positions downstream of the objective. Forcing was applied in the band $0.3 \leqslant St \leqslant 0.45$ and control was carried out with the $K_{r}$ kernel with $d$ as input. The difference between uncontrolled and controlled jets is clear in the spectra and in rms levels up to $x/D=7$. Close to the objective position, rms reduction is restricted to radial positions close to the jet axis, and an amplification effect occurs at $0.25 \leqslant r/D \leqslant 0.5$. This undesired phenomenon is associated with high-frequency content close to the objective position, as can be seen in the spectra of Figure \ref{control_res}. However, these scales gradually lose energy as they convect downstream and the correct trend of reduction is obtained across the shear-layer. The results show that, even though the control strategy has a localised character insofar it is formulated to achieve its objective at a specific position within the jet, it produces reductions of wavepacket amplitudes throughout the jet. 

To conclude, we have demonstrated that wave cancellation of stochastic disturbances is possible in a turbulent jet. The attenuation of the disturbances is found to occur both at the objective position and over an extended streamwise region downstream of this. Working with a forced jet was a necessary first step for the longer-term objective of closed-loop control of an unforced jet. The latter problem requires a more refined choice of sensors and their positioning in order to avoid azimuthal aliasing issues, for instance. The performance of the control is shown to be largely underpinned by coherence between sensor and objective, on one hand, and actuator and objective on the other. This suggests that these coherence metrics can be used to guide optimisation of sensor and actuator placement for the control of unforced jets.

I. A. M. acknowledges support from the Science Without Borders program through the CNPq grant number 200676/2015-6. E. M. acknowledges support from CAPES grant 88881.190271/2018-01. P. J, A. V. G. C. and F. J. S. acknowledge support from the CAPES Science Without Borders project no. A073/2013. The authors wish to thank Redouane Kari and Damien Eysseric for their invaluable work during the experimental campaign.

\bibliographystyle{unsrtnat}
\bibliography{bibfile}

\end{document}